\documentclass[%
 aip,
 amsmath,amssymb,
 reprint,%
]{revtex4-1}

\usepackage{graphicx}% Include figure files
\usepackage{dcolumn}% Align table columns on decimal point
\usepackage{bm}% bold math
\usepackage[dvipsnames]{xcolor}
\usepackage[utf8]{inputenc}
\usepackage[T1]{fontenc}
\usepackage{mathptmx}
\usepackage{etoolbox}

\makeatletter
\def\@email#1#2{%
 \endgroup
 \patchcmd{\titleblock@produce}
  {\frontmatter@RRAPformat}
  {\frontmatter@RRAPformat{\produce@RRAP{*#1\href{mailto:#2}{#2}}}\frontmatter@RRAPformat}
  {}{}
}%
\makeatother
\begin{document}
\setlength\belowcaptionskip{-5pt}
\preprint{AIP/123-QED}

\title[Room-temperature, continuous-wave Terahertz generation 
in free-space with an intersubband mid-infrared photomixer]{Room-temperature, continuous-wave Terahertz generation 
in free-space with an intersubband mid-infrared photomixer}
\author{Q. Lin}
\author{J-F. Lampin}
\affiliation{Institute of Electronics, Microelectronics and Nanotechnology, CNRS, Univ. Lille,  Univ. Polytechnique Hauts-de-France, UMR 8520, F-59000 Lille, France
}%
\author{G. Ducournau}
\affiliation{Institute of Electronics, Microelectronics and Nanotechnology, CNRS, Univ. Lille,  Univ. Polytechnique Hauts-de-France, UMR 8520, F-59000 Lille, France
}%
\author{S. Lepillet}
\affiliation{Institute of Electronics, Microelectronics and Nanotechnology, CNRS, Univ. Lille,  Univ. Polytechnique Hauts-de-France, UMR 8520, F-59000 Lille, France
}%
\author{H. Li}
\affiliation{%
Key Laboratory of Terahertz Solid State Technology, Shanghai Institute of Microsystem and Information Technology, Chinese Academy of Sciences, 865 Changning Road, Shanghai 200050, China%\\This line break forced% with \\
}%
\author{E. Peytavit}
\affiliation{Institute of Electronics, Microelectronics and Nanotechnology, CNRS, Univ. Lille,  Univ. Polytechnique Hauts-de-France, UMR 8520, F-59000 Lille, France
}%
\author{S. Barbieri}
\affiliation{Institute of Electronics, Microelectronics and Nanotechnology, CNRS, Univ. Lille,  Univ. Polytechnique Hauts-de-France, UMR 8520, F-59000 Lille, France
}%
\email{stefano.barbieri@iemn.fr} %% email address is required; see note below about the corresponding author designation

\date{\today}% It is always \today, today,
             %  but any date may be explicitly specified

\maketitle

\begin{quotation}
We demonstrate a Terahertz photomixer pumped by mid-infrared photons at 10$\mu$m wavelength. The device is based on a dual-antenna architecture, in which a two-dimensional array of mid-infrared patch-antenna resonators is connected to the electrodes of a log-spiral Terahertz antenna.
By exploiting intersubband absorption inside an AlGaAs-GaAs multi-quantum-well heterostructure, a photocurrent is coherently generated at the difference frequency between two quantum cascade lasers. The photocurrent drives the antenna electrodes allowing tunable, continuous-wave generation in free-space up to 1 Terahertz at room temperature. By experimentally studying the photomixer frequency response and dark impedance, we demonstrate that the observed power roll-off at high-frequency is limited by the combined effect of the photoexcited carrier's intrinsic lifetime and of the device $RC$ time constant. In the spectral range investigated we obtain, in continuous-wave,  mid-infrared $\rightarrow$ Terahertz conversion efficiencies exceeding by orders of magnitude those of unipolar devices exploiting $\chi^{(2)}$ processes.
\end{quotation}
\section{INTRODUCTION}
 In a continuous-wave photomixer (PM), Terahertz (THz) light is radiated by the $ac$ component of the photocurrent generated by beating a pair of slightly detuned diode lasers, focused on a low-carrier lifetime photoconductor (e.g. low-temperature grown GaAs, InGaAs:Fe, etc.) or an ultrafast photodiode~\cite{Peta-2021}. Contrary to non-linear optics generation, this process is not limited by Manley-Rowe relations, typically allowing to reach higher conversion efficiencies~\cite{Preu-2011,Ohara-2023,Maes-2023,Peta-2011}. This is the reason why PMs have become the most popular photonic devices for continuous-wave (CW) THz generation, ever since, back in 1995, Brown \textit{et al.} identified GaAs grown at low-temperature as an ideal material thanks to its sub-ps electron-hole recombination lifetime, high-mobility and high breakdown field~\cite{Preu-2011,Brown-95-1,Brown-95-2}.

With the exception of THz quantum cascade lasers (QCLs), so far THz emission based on intersubband (ISB) transitions in unipolar devices has been obtained through $\chi^{(2)}$ difference frequency generation (DFG) in systems of coupled quantum wells (QWs)~\cite{Dupont-2006}. Since the pioneering work by Sirtori \textit{et al.}, the exploitation of this technique has culminated with the demonstration, in 2007, of intra-cavity  
	\begin{figure}[h]
 %\vspace*{-1.5cm}
	\includegraphics[width=0.47\textwidth]{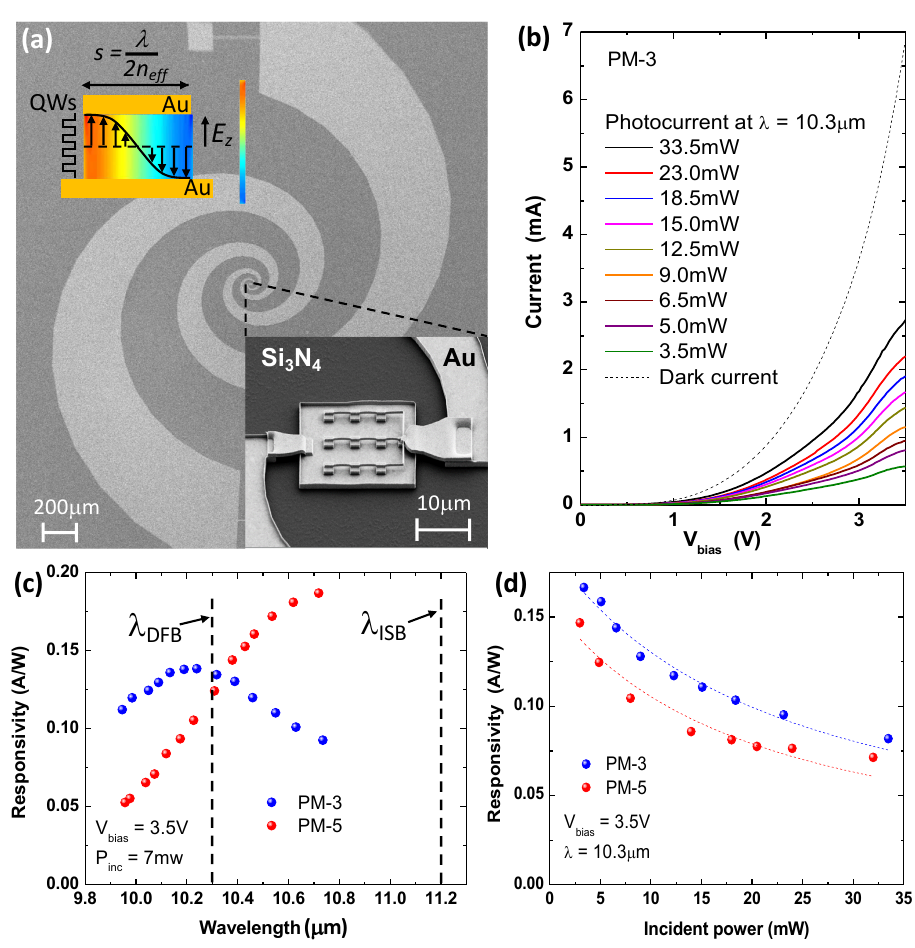}
    \caption{\label{fig:Design} (a) Main panel. SEM photograph of PM-3 showing the arms of the log-spiral antenna. Bottom Inset. SEM photograph of the 3$\times$3 patch-antenna matrix, electrically connected to the antenna arms by metallic air bridges. Top Inset. Schematic cross-section of a patch-antenna resonator. The amplitude of the electric field in the $z$ direction associated to the the TM$_{10}$ mode is shown in color scale. $n_{eff}$ is the mode effective index. (b) Dark current (dashed line) and photocurrent (i.e. current under illumination minus dark current - solid lines) $vs$ bias voltage characteristics of PM-3, for different incident powers. The device was illuminated by a QCL emitting radiation at $\lambda = 10.3 \mu$m, linearly polarized perpendicularly to the wires connecting the patches. (c) Responsivity $vs$ wavelength characteristics of PM-3 and PM-5 measured under an applied bias of 3.5V and an incident power of 7mW.   The vertical dashed lines indicate the emission wavelength of the DFB QCL used in this work, and the peak wavelength of the ISB transition. (d)  Responsivity $vs$ incident power characteristics of PM-3 and PM-5 measured at $\lambda = 10.3 \mu$m, under an applied bias of 3.5V. The dashed lines are the results of the fits (see text).}
	\end{figure}
DFG in mid-infrared QCLs~\cite{Sirtori-94,Belkin-2007,Fujita-2018,Raz-2016}.
In this work we show instead that THz emission in free-space can also be obtained from microscopic current oscillations, as a result of direct photon absorption between two electronic subbands in a GaAs-AlGaAs multi-quantum-well heterostructure.  By exploiting an original dual-antenna architecture, and by taking advantage of the fact that electron recombination ruled by optical-phonon emission takes place on a timescale of the order of 1 ps, we demonstrate generation up to 1THz at room temperature. 
 
\section{RESULTS}\label{Results} 
SEM pictures of the devices fabricated through e-beam litography are shown in Fig.~\ref{fig:Design}(a) (see Methods for fabrication details). They consist of a three-turn Ti/Au log-spiral THz antenna, of $72\Omega$ nominal impedance, with, at its center, the MIR photon absorbing area. The latter is made of a 2D-array of patch-antenna resonators connected in parallel by suspended metal wires, and sharing a common ground plane (Fig.~\ref{fig:Design}(a), bottom inset). Both the ground plane and antenna arms sit on top of a 100-nm-thick Si$_{3}$N$_{4}$ layer deposited over a high-resistivity (HR) Silicon substrate. The patches top contacts and ground plane are respectively connected to the top and bottom arms of the spiral antenna by metallic air bridges. The top of each resonator is a square metallic patch of side $s \sim 2\mu$m, with, underneath, a semiconductor heterostructure containing seven, 6.5nm GaAs quantum-wells (QW) separated by 40nm Al$_{0.2}$Ga$_{0.8}$As barriers. As shown in the top inset of Fig.~\ref{fig:Design}(a), the patch side sets the wavelength, $\lambda_{res}\sim10 \mu$m, of the fundamental TM$_{10}$ mode of the resonators. While oscillating in the horizontal plane, hence with the electric field $E_{z}$ perpendicular to the semiconductor layers, as required by ISB selection rules, this mode can be excited by a MIR plane wave at normal incidence, with a polarization component along the y-axis, which is particularly convenient to pump the PM~\cite{Rodriguez2022}. ISB absorption, ideally coincident with $\lambda_{res}$, takes place from the QWs bound-state up to a quasi-bound state close to the top of the barriers, generating a photocurrent perpendicular to the layers when the PM active region is $dc$ biased through the antenna arms~\cite{Liu2007}. In particular, when simultaneously illuminating the PM with two CW MIR lasers, a photocurrent with an $ac$ component oscillating at their difference frequency is generated. This behaves as a source of electromagnetic radiation that is finally emitted in free-space via the spiral antenna. By taking advantage of the fact that both the patch resonators mode and ISB transition are spectrally broad, continuously tunable radiation can be obtained, from microwaves up to potentially a few THz, by changing the wavelength of one laser~\cite{Hakl2021}.

Two sets of PMs, labelled PM-3 and PM-5, were fabricated for this work, based on patch matrices of 3$\times$3 and 5$\times$5 squares of side $s = 1.8\mu$m and $1.9\mu$m respectively, separated by a period $p = 5\mu$m. Through reflectivity measurements on larger matrices (not shown) using a MIR microscope coupled to a FTIR spectrometer, we verified experimentally that, for incident light polarized perpendicularly to the the wires connecting the patches, this specific value of $p$ allows to obtain approximately $\sim$90$\%$ radiation absorption at $\lambda_{res} \simeq 10.2 \mu$m, for $s = 1.8\mu$m, and $\lambda_{res} \simeq 10.7 \mu$m for $s = 1.9\mu$m . This is the case when the focused MIR source has a spotsize much smaller that the patch matrix, and means that a single-patch absorbs nearly all the radiation incident on a surface equal to approximately 6 times its physical area. This property, originating from the fact that the patch resonators behave as antennas~\cite{Rodriguez2022}, is important for this work since it allows to minimize the PMs capacitance without compromising the radiation collection efficiency ~\cite{Hakl2021,Lin-2023}.

In Fig.~\ref{fig:Design}(b) are displayed the photocurrent vs bias characteristics of PM-3 under different illumination powers from a distributed feedback (DFB) QCL emitting at 10.3$\mu$m, and linearly polarized perpendicularly to the connecting wires. These measurements, like all the data presented in this work, were recorded at 300K. The dashed black line is the PM dark current. Its non-linear behaviour stems from the presence of Schottky barriers at the metal semiconductor interfaces. In Fig.~\ref{fig:Design}(c) we report the spectral dependence of the devices responsivities at $V_{bias}=3.5$V, measured in the range 9.8-10.7$\mu$m with an external cavity (EC) QCL. Their different behavior is a result of the different $\lambda_{res}$. These are both blue-shifted relative to the peak wavelength of the ISB transition, $\lambda_{ISB}\sim 11.2\mu$m, that was measured independently (see Ref.[15]). Moreover, for PM3, the collection area of the patch matrix ($\sim15 \times 15\mu$m$^{2}$) is  smaller than the laser spot ($\sim25 \mu$m diameter), reducing the absorption. The resulting device responsivities ($R_{esp}$) almost match at $\sim 10.3\mu$m, corresponding to the DBF QCL emission wavelength. 

In Fig.~\ref{fig:Design}(d) the responsivities for different incident powers are reported at $V_{bias}=3.5$V ($\lambda = 10.3\mu$m), showing a clear drop with increasing power. Responsivity saturation with incident power in QW MIR photo-detectors has been observed in the past at low temperatures and explained in terms of screening of the applied electric field induced by the photo-excited carriers. ~\cite{Ershov-1995,Ershov-1997,Mermelstein-1997,Liu2007} A necessary condition for this to happen is that the photocurrent is at least of the same magnitude as the dark current. From Fig.1(d) this is not the case for our devices operating at 300K, where saturation is already effective at a few mW incident power, i.e. with a photocurrent that is less than 10$\%$ of the dark current (Fig.1(b)). Instead, we attribute the observed responsivity reduction to optical saturation of the ISB transition. From the fits using the function $R_{esp} = R_{0}$/(1 + $P_{inc}$/$P_{sat}$), where $P_{inc}$ is the incident power, while $R_{0}$ and $P_{sat}$ are used as fitting parameters, we obtain a saturation power $P_{sat} \sim 20$mW, i.e., considering the laser spot-size, an incident saturation intensity of $\sim 4$kW/cm$^{2}$. This value is in agreement with our calculations (see Ref.[16] for a detailed derivation) and stems from the presence of the patch resonators, effectively reducing the saturation intensity by over two orders of magnitude compared to that of a bare ISB transition~\cite{Vodopyanov_1997,Lin-2023,Tymchenko2017,Jeannin2020-2,Jeannin2023}.

The schematic of the experimental setup used to characterize free-space THz emission from the PMs is displayed in the inset of Fig.~\ref{fig:Resp}. The DFB and EC QCLs, each generating $\sim 20$mW of incident power and linearly polarized perpendicularly to the patch wires, are collinearly focused on the patch matrix biased at 3.5V via the antenna arms. The electromagnetic radiation emitted by the spiral antenna can be tuned from microwaves to THz by changing the frequency of the EC QCL, and is collimated with the help of a hyper-hemispherical Si lens positioned on the backside of the $500\mu$m-thick HR Si substrate. 
	\begin{figure}[h!]
%\vspace*{-1.5cm}	
 \includegraphics[width=0.47\textwidth]{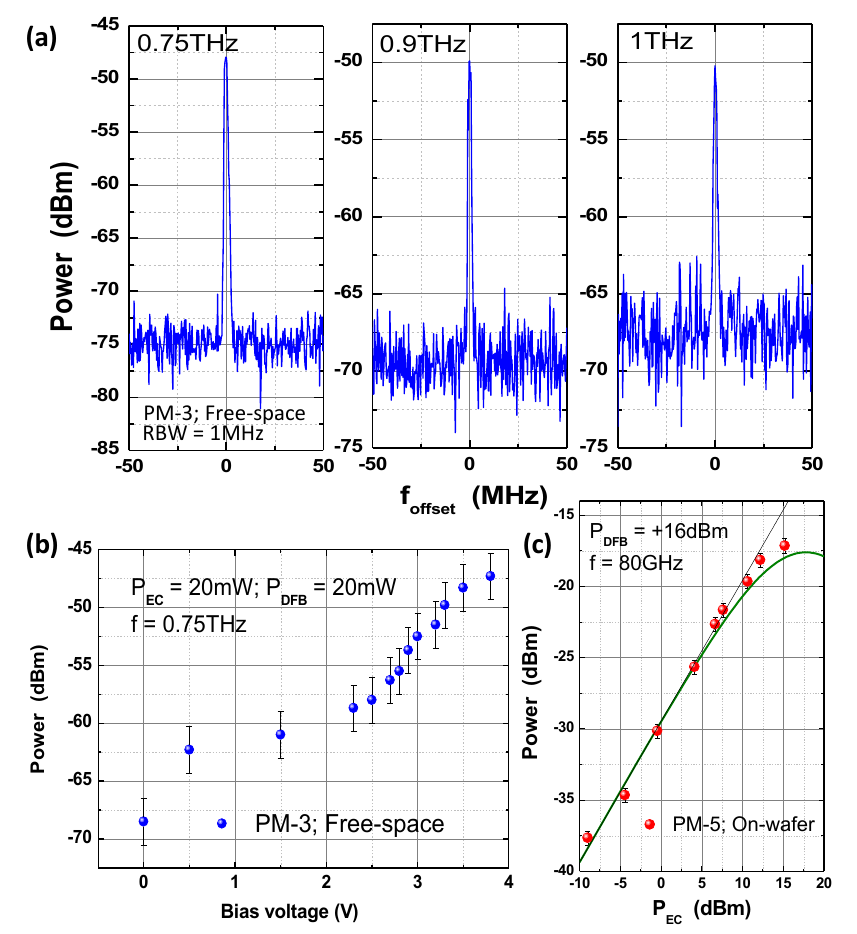}
    \caption{\label{fig:Mixer} (a) Heterodyne beatnotes measured at the IF output of the HM ($\lambda_{DFB}=10.3\mu$m) for $P_{DFB} = P_{EC} \simeq 20$mW. For all spectra the resolution bandwidth is 1MHz. The mixer (VDI, model WR1.0SAX-M) covers the 0.75-1.1THz band, with a 144 multiplication factor. The resulting local oscillator frequency range is 5.21-7.65GHz. The beatnotes spectra are corrected by the HM mixer conversion factors of 10.5dB, 16.2dB, and 17.2dB respectively at 0.75THz, 0.9THz and 1THz. (b) Bias dependence of the detected power from PM-3 at f = 750GHz. (c) Detected power at f = 80GHz $vs$ $P_{EC}$ while the PM is illuminated by the DFB QCL with $P_{DFB} =40$mW (+16dBm). The device is based on a 5$\times$5 patch matrix, nominally identical to that of PM-5, connected to an integrated 50$\Omega$ coplanar line and probed by a coplanar probe.}
	\end{figure}
A f-2 Zeonex lens is finally used to focus the THz radiation on a set of harmonic mixers (HMs), allowing to measure the device emission in the bands 75-110GHz, 140-220GHz, 250-400GHz (RPG ZRX Series \cite{RPG}), and 0.75-1.05THz (VDI Mini-SAX Series \cite{VDI}). 

Examples of down-converted spectra at 0.75THz, 0.9THz and 1.05THz, obtained from PM-3 under an applied bias of 3.5V, are reported in Fig.~\ref{fig:Mixer}(a). As detailed in the figure caption the spectra are corrected by the HM conversion factors, which take into account the intrinsic conversion loss of the mixer as well as the amplifiers gain (for all the HMs used in this work, we verified experimentally the conversion factors given by the manufacturer with the help of a calibrated power meter and a set of calibrated sub-mm electronic sources). The SNR results from the combination of a resolution bandwidth of 1 MHz (imposed by the linewidth of the lasers used to pump the photomixer), of the receiver conversion losses, and of an additional 3dB attenuation due to the fact that the circularly polarized THz radiation emitted by the spiral antenna is fed into the HMs through a rectangular single-mode waveguide ~\cite{Ravaro-2011}. At 1THz we obtain a SNR of $\sim 17$dB, and the resulting MIR $\rightarrow$ THz conversion efficiency, defined as the ratio between the emitted THz power and the product of the MIR pumps, is of $\sim 6\times10^{-5}$W$^{-1}$. 

\begin{figure}
%\vspace*{-1.5cm}	
 \includegraphics[width=0.47\textwidth]{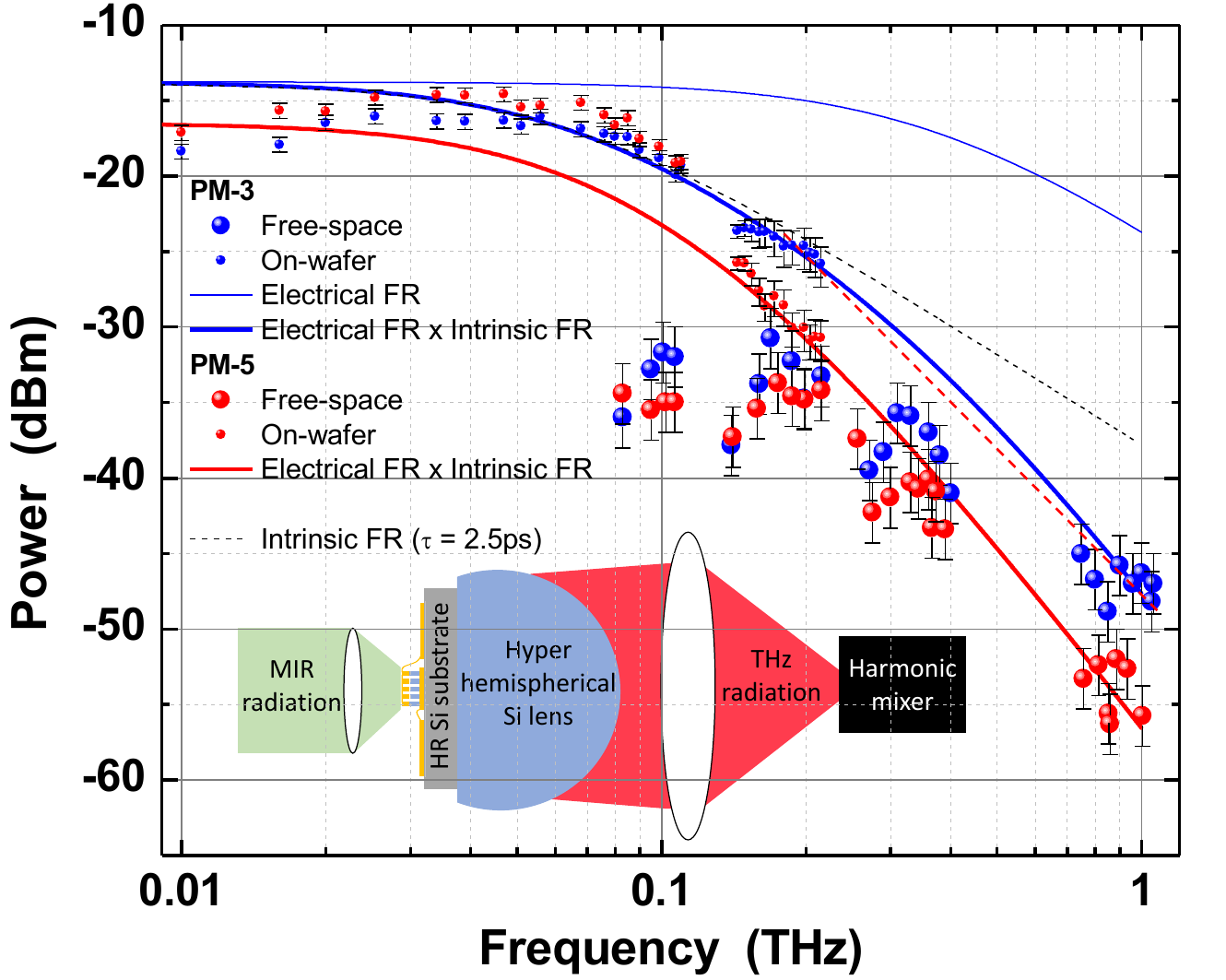}
    \caption{\label{fig:Resp} Detected power $vs$ frequency for PM-5 (red) and PM-3 (blue). Small circles are related to devices with an integrated $50\Omega$ coplanar access, that we characterized on-wafer with the help of coplanar probes. Large circles are related to devices with a spiral antenna (Fig.~\ref{fig:Design}(a)). The devices with coplanar access were biased with the help of a bias-T, and the experimental data points are corrected by the losses of the bias-T and coplanar probes, that we measured with a VNA. The data points related to the PMs with spiral antenna are measured in free-space, and are corrected by the HMs conversion factors, and by the 3dB attenuation due to the polarization mismatch between the spiral antenna and the HMs input rectangular waveguide (see main text). The thick blue and red solid lines are the PMs FRs obtained from the product of the electrical FR (computed from the small-signal equivalent circuit model shown in the inset of Fig.~\ref{fig:Impedance}) and the intrinsic FR, with a carriers lifetime of 2.5ps. The latter is shown by the black dashed line. The thin solid blue line shows the computed electrical FR of PM-3. The red dashed line has a slope of 35dB/decade. Inset. Schematic drawing (not to scale) of the experimental setup used to characterise the PMs in free-space. The Si hyper-hemispherical lens has a diameter of 5mm and is in direct contact with the backside of the 500$\mu$m-thick HR Si substrate. The two white lenses represent a f-1, HR-coated MIR chalcogenide lens and a f-2, THz Zeonex lens.}
	\end{figure}

In Fig.~\ref{fig:Mixer}(b) we report the bias dependence of the detected power at 0.75THz, showing a clear saturation as the bias gets close to 4V. We deliberately avoided raising the bias beyond 3.8V to minimize the risks of damaging the PMs. 
	\begin{figure}[h!]
%\vspace*{-1.5cm}	
 \includegraphics[width=0.47\textwidth]{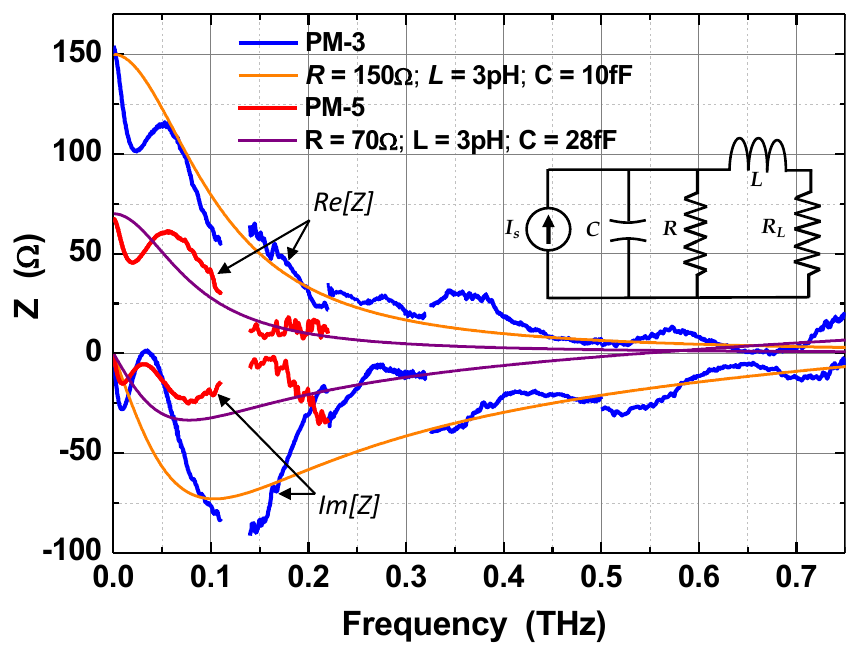}
    \caption{\label{fig:Impedance} Real and imaginary parts of the dark impedance of the patch matrix of PM-3 (blue lines) and PM-5 (red lines). Orange solid-lines: fit of PM-3  dark impedance (real and imaginary) using the small-signal equivalent circuit model reported in the inset, with $C = 10$fF and $R$, $L$ as fitting parameters. The resulting values of $R$ and $L$ are given in the legend. Purple solid lines: dark impedance of PM-5 computed using the values of $C$, $R$ and $L$  given in the legend. Inset. Small-signal equivalent circuit model of the patch matrix. $I_{s}$ is a constant current source proportional to the generated photocurrent (see text). $R_{L}$ is the load impedance, equal to $50\Omega$ for the coplanar line, or $72\Omega$ for the spiral antenna.}
	\end{figure}
The red circles in Fig.~\ref{fig:Mixer}(c) correspond to the measured RF power at 85GHz, and were obtained by keeping the PM under constant illumination from the DFB QCL with an incident power $P_{DFB} =40$ mW (+16dBm), while increasing the power from the EC QCL, $P_{EC}$, from = 0.125 mW (-9dBm) to 33 mW (+15dBm). For this measurement, we used a device based on a 5$\times$5 patch matrix, nominally identical to that of PM-5, where the spiral antenna was replaced by a 50$\Omega$ tapered coplanar line (see Fig.~\ref{Coplanar}), allowing power measurements directly on-wafer with the help of coplanar probes and a calibrated microwave power meter. As shown in Fig.~\ref{fig:Mixer}(c), up to $P_{EC} \simeq +7$dBm (5mW), the experimental points follow a unit slope (solid black line), which is expected for a constant optical responsivity, since the generated microwave power is proportional to $P_{EC} \times P_{DFB}$. At higher powers we find instead a clear deviation that can be explained on the basis of the observed responsivity saturation (Fig.~\ref{fig:Design}(d))). This is shown by the solid green curve, obtained from the expression of the generated microwave power where, for the responsivity, we used the fit in Fig.~\ref{fig:Design}(d) (red dashed line).

The measured free-space THz power $vs$ frequency, corrected by the 3dB attenuation due to the polarization mismatch between the spiral antenna and the HMs input waveguide, is reported in Fig.~\ref{fig:Resp} for PM-3 and PM-5 (blue and red large circles).
Two additional PMs with an integrated $50 \Omega$ coplanar line were used to complete the frequency response (FR) in the microwave frequency range using two sets of coplanar probes and calibrated power meters (0-110GHz and 140-220GHz bands - small blue and red circles). One of them is the device used for Fig.~\ref{fig:Mixer}(c), while the second is based on a 3$\times$3 patch matrix nominally identical to that of PM-3 (given the experimental errors in Fig.~\ref{fig:Resp} the difference between the nominal impedance of the coplanar line and that of the spiral antenna is completely negligible). For all data points shown in Fig.~\ref{fig:Resp} the PMs were operated with an applied bias of 3.5V, and approximately 20mW incident power from each QCL, resulting in a total $dc$ photocurrent of $\sim 3$mA. 
For both PM-3 and PM-5 we find a 3dB cutoff close to 100GHz. As expected, in the 75-110GHz  and 140-220GHz bands, the amount of THz power collected from the coplanar line is much larger compared to free-space. Indeed, at low frequencies, despite the presence of the Si hyper-hemispherical lens, emission in free space is highly divergent, severely reducing the collection of radiation by the Zeonex lens (moreover, in this frequency range the impedance of the antenna cannot anymore be considered constant). By moving the lens laterally, we verified experimentally that in the range 0.75-1.05THz, emission is instead much more directional, allowing for the measurement of the total emitted power. As a result, only the points belonging to this frequency band, together with those relative to the devices with coplanar access, are relevant for the evaluation of the PMs FR. The latter is the product of two terms: (i) the intrinsic FR, proportional to $1/(1+(2\pi f\tau)^2)$, where $\tau$ is the electron's recombination (or capture) time; (ii) the electrical FR that depends on the complex impedance of the patch antenna matrix \cite{Brown-95-1,Preu-2011,Hakl2021}. From Fig.~\ref{fig:Resp}, the measured $\sim 100$GHz 3dB cutoff sets an upper limit of $\sim 2\div 3$ps for $\tau$. On the other hand it appears that, regardless of the value of $\tau$, the intrinsic contribution alone (dashed black line, for $\tau = 2.5$ps), yielding a $20$dB/decade high-frequency roll-off , is not sufficient to explain the $\sim 35$dB/decade roll-off observed experimentally (dashed red line). Therefore the PM impedance must be taken into account to model the PMs electrical FR. In principle the latter should be measured under illumination, which, presently, is technically out of reach. However, for the illumination power used ($\simeq 20$mW+$20$mW), the total PM current is still dominated by the dark component (see Fig.~\ref{fig:Design}(b)). Therefore, contrary to interband PMs, we can assume that the PM impedance under illumination can be reasonably well approximated by its value in the dark. 

To obtain the dark impedance of the 3$\times$3 patch matrix we measured on-wafer its $S_{11}$ parameter in the range 0-750GHz using a series of Vector Network Analyser (VNA) extension modules with coplanar probe access (from Rohde \& Schwarz and from VDI in the 500-750GHz band). The real and imaginary part of the dark impedance under an applied bias of 3.5V, obtained after de-embedding the coplanar line, are reported in Fig.~\ref{fig:Impedance}. The orange solid-lines are the result of a fit using the small-signal equivalent circuit model reported in the inset. Here $R$ and $L$, representing respectively the PM resistance and inductance (introduced by the suspended metal wires), are the fitting parameters, while $C = 10$fF corresponds to the computed static capacitance of the patch matrix. As shown in the Figure, with $R \simeq 150\Omega$ (corresponding to the device resistance at 0 Hz) and $L\simeq 3$pH, our simplified model reproduces reasonably well the impedance general behavior, namely the pronounced drop at high frequencies. The solid red lines in Fig.~\ref{fig:Impedance} are the real and imaginary dark impedance of the 5$\times$5 patch matrix, measured up to 220GHz, since in this case we did not fabricate devices with a coplanar access adapted to the full set of VNA probes. To model the impedance we used the small-signal equivalent circuit with $C = 28$fF, corresponding to the static capacitance of the set of 25 patch resonators, $R = 70\Omega$, equal to the measured static resistance (see Fig.~\ref{fig:Impedance}) and $L = 3$pH, i.e. identical to that of the 3$\times$3 matrix (we note that $L$ affects only the imaginary part of the impedance, and that up to $\sim 10$pH it does not change significantly the device FR). The resulting impedance for the 5$\times$5 matrix is shown by the purple solid lines in Fig.~\ref{fig:Impedance}.

The thin solid blue curve in Fig.~\ref{fig:Resp} is the electrical FR of PM-3, obtained from the equivalent circuit by computing the average electrical power dissipated in the antenna resistance $R_{L} = 72\Omega$, considering a constant current source $I_{s}=1.6$mA~\cite{Lin-2023}. The thick solid blue curve is obtained by multiplying the electrical FR by $1/(1+(2\pi f\tau)^2)$ with $\tau=2.5$ps (dashed black line), showing a good agreement with the experimental points (small blue circles, and large blue circles in the range 0.75-1.05THz). The solid red curve is the corresponding FR for PM-5, still computed with $\tau=2.5$ps,  showing that close to 1THz, the power drop with respect to PM-3 can be explained by the increase in the device capacitance.  
\section{DISCUSSION} 

\begin{figure}
%\vspace*{-1.5cm}	
 \includegraphics[width=0.47\textwidth]{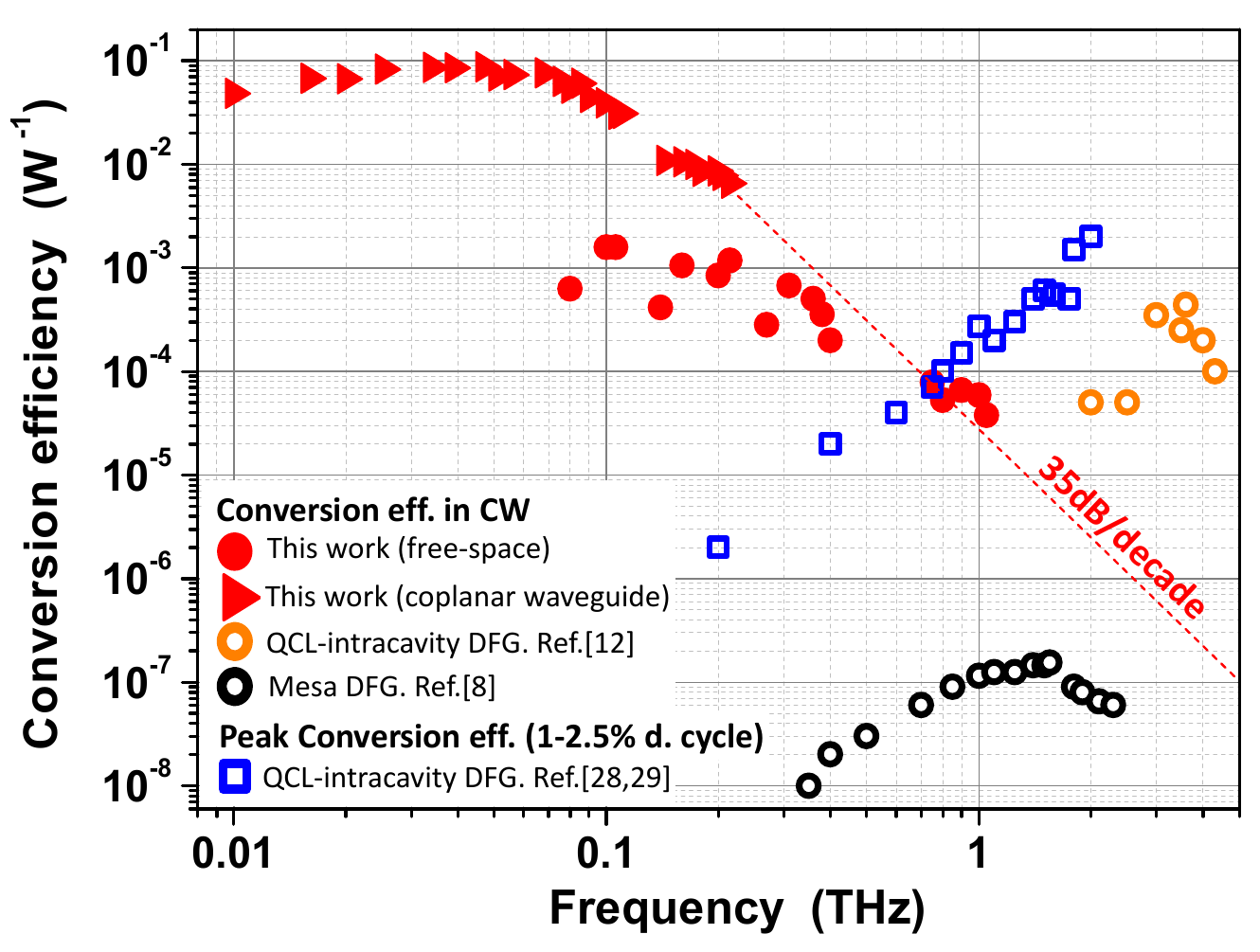}
    \caption{\label{Efficiency} Room temperature MIR$\rightarrow$THz conversion efficiency for this work (red triangles and circles, CW operation); intra-cavity DFG (orange open circles, CW operation~\cite{Raz-2016}; blue open squares, pulsed operation with 1-2.5$\%$ duty cycle~\cite{Fujita-2022,Hayashi-2024}); DFG in an optically-pumped mesa~\cite{Dupont-2006} (black open circles, CW operation). The red dashed line has a slope of 35dB/decade.}
	\end{figure}

\begin{figure}[h!]
%\vspace*{-1.5cm}	
 \includegraphics[width=0.37\textwidth]{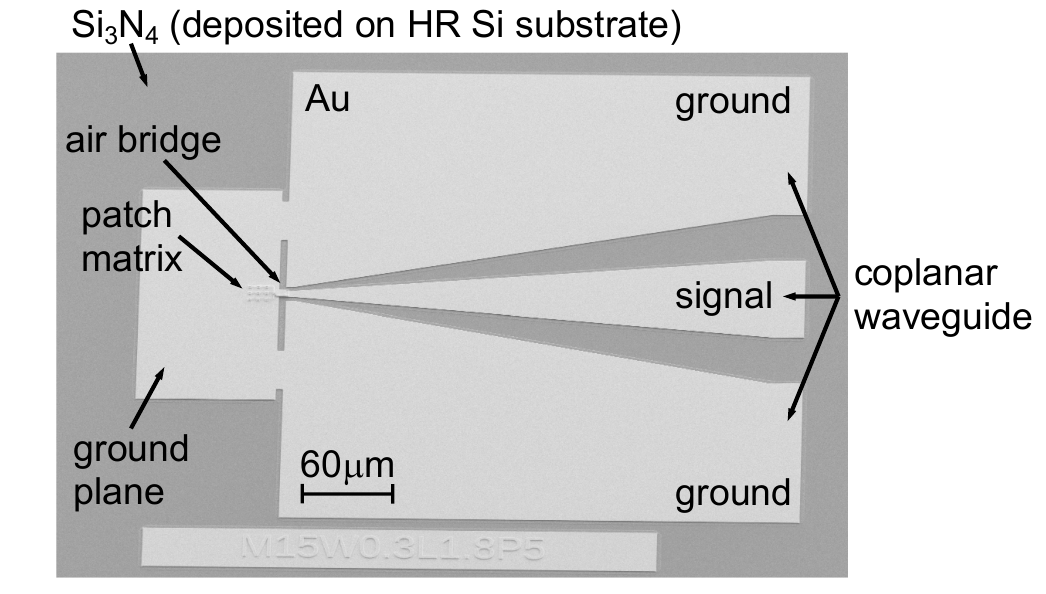}
    \caption{\label{Coplanar} SEM photograph of PM-3 with coplanar waveguide access.}
	\end{figure}

As pointed out in the introduction, so far THz generation based on optically-pumped ISB transitions has exploited $\chi^{(2)}$ DFG processes. Instead, in this work we have shown unequivocally and for the first time that THz waves can also be radiated in free-space by current oscillations induced by direct photon absorption between two electronic subbands. Since for both DFG and photomixing, the generated THz power is proportional to the product of the pumps, it is interesting to compare the two techniques by computing the MIR$\rightarrow$THz conversion efficiency (CE), in units of W$^{-1}$, given by the relation $P_{THz}/(P_{MIR-1}\times P_{MIR-2})$, where $P_{THz}$ is the generated THz power, and $P_{MIR-1}$, $P_{MIR-2}$ the MIR pump powers . This is shown in Fig.~\ref{Efficiency} where we report (red triangles and circles) the CEs obtained from the data of Fig.~\ref{fig:Resp}, together with what are, to our knowledge, the best experimental results from QCL intra-cavity DFG at room-temperature (orange circles and blue squares). In the figure the open orange circles are obtained in CW~\cite{Raz-2016}, while the blue open squares represent the peak conversion efficiencies obtained in pulsed operation with a duty cycle of 1-2.5$\%$ (we could not find experimental data in CW below ~$\sim$ 2THz)~\cite{Hayashi-2020,Fujita-2022,Hayashi-2024}. Finally, the open black circles are the CEs obtained by Dupont \textit{et al.} from a mesa-like (i.e. without micro-cavity) AlGaAs-GaAs hetero-structure, consisting of 100-repeats of two coupled QWs, externally pumped by two CW CO$_{2}$ lasers at $\lambda \sim 10\mu$m~\cite{Dupont-2006}. Regardless of the technique, de DFG CE drops at low frequencies, a result due to the $\omega^{2}$-dependence of the generated power, but also to increasing free carrier absorption, decreasing optical non-linearity, and diffraction effects~\cite{Sirtori-94,Dupont-2006,Fujita-2019,Fujita-2022}. For the PMs of this work the behavior is instead exactly the opposite, reflecting the completely different THz generation process, with, as demonstrated, a high-frequency roll-off deriving from both the devices electrical FR and carriers capture. The resulting crossing point between the two approaches is around 0.7-0.8THz, although it should be noted that the DFG data (blue open squares) correspond to peak CEs.

Beyond ISB DFG generation, the performance of the demonstrated devices should be benchmarked against that of state-of-the-art near-infrared CW photomixers based on GaAs and InGaAs, or InGaAs/InP uni-traveling-carrier photodiodes (UTC). Although the gap in emitted power/CE is presently of several orders of magnitude, with, for instance, record CEs of a few W$^{-1}$ at 270GHz and of $0.1$ W$^{-1}$ at 250GHz, respectively for UTC and GaAs-based photomixers ,~\cite{Ohara-2023,Peta-2011} however we believe that the results presented in this work constitute a relevant proof of principle, paving the way to further developments and improvements. Theoretically, for equal optical power, the generated photocurrent should scale as the ratio between the wavelengths of the pump photons, which motivates future investigations of MIR photomixers. A detailed discussion on how to optimise the demonstrated PMs is beyond the scope of this work, however the results obtained indicate some possible routes. 
First, the generated photocurrent can be increased by maximizing the PM quantum efficiency, namely by adjusting $\lambda_{res}$ to match $\lambda_{ISB}$ (see Fig.~\ref{fig:Design}(c)). We estimate that in this way the latter could be raised from the actual $\sim15\%$ (for PM-5) to $\sim30\%$, thus increasing by a factor of $\sim 4$ the actual CE. On the other hand, as shown in Fig.~\ref{fig:Mixer}(c), the maximum generated THz power is presently limited by saturation. In this respect, a first crucial point to address experimentally would be to unambiguously establish the origin of the observed power saturation, e.g. by illuminating the devices with short optical pulses or by MIR pump-probe spectroscopy. Assuming, as we believe (see Section ~\ref{Results} and Fig.~\ref{fig:Design}(d)) that the origin is mainly optical saturation of the ISB transition (and not thermal or space-charge effects~\cite{Ershov-1995,Ershov-1997,Mermelstein-1997,Liu2007}), then the saturation power is proportional to $n_{s} \times N_{QW}/\tau$, where $n_{s}$ is the charge density in the QWs, and $N_{QW}$ the number of QWs ~\cite{Jeannin2020-2,Lin-2023}. The saturation power could then be raised by increasing $n_{s}$ and/or $N_{QW}$, while keeping the same length of the active region in order to maintain the electron’s transit time across the PM heterostructure, $\tau_{tr}$, unaltered. To this end we expect that it should be possible to reduce the thickness of the AlGaAs barriers, from the actual 40nm, down to about 20nm, still preventing significant inter-well tunneling~\cite{Liu2007}. Reducing the barrier width could even possibly lead to a reduction of the electron's capture time, thus widening the PM's bandwidth. Another solution to explore is to raise the generated photocurrent by lowering the device temperature, leading to an increased mobility (i.e. a smaller $\tau_{tr}$) ~\cite{Palaferri2018,Hakl2021}. Finally, in the case where broad frequency tuning is not needed, we should also mention the possibility to replace the broadband spiral antenna used in this work with one based on a resonant dipole, which has been shown to increase the THz emitted power by up to a factor of $\sim 5$~\cite{McIntosh-1996,Gregory-2004}. 

\section{METHODS} 
\subsection{Device Fabrication}
A 100nm-thick, lattice-matched Ga$_{0.51}$In$_{0.49}$P etch-stop layer followed by the Al$_{0.2}$Ga$_{0.8}$As/GaAs heterostructure is grown by MBE on top of a semi-insulating GaAs substrate. The heterostructure is sandwiched between 50 and 100nm-thick top and bottom n-doped contact layers with concentrations $4 \times 10^{18}$cm$^{-3}$ and $3 \times 10^{18}$cm$^{-3}$, and consists of seven, 6.5nm-thick GaAs QWs with the central 5.3nm $n$-doped at $6 \times 10^{17}$cm$^{-3}$, separated by 40nm-thick, undoped Al$_{0.2}$Ga$_{0.8}$As barriers. 

To fabricate the PMs, the epi-layer is initially transferred onto a 2”-diameter HR Si wafer using Au–Au thermo-compression bonding, with a 100-nm-thick Si$_{3}$N$_{4}$ layer deposited on the Si wafer beforehand using plasma-enhanced chemical vapor deposition (PECVD). The fabrication begins with the removal of the GaAs substrate and the etch-stop layer through wet etching. Next, 2D-patch arrays of top Ti/Au (8nm/300nm) Schottky contacts are fabricated by e-beam lithography, followed by e-beam evaporation and lift-off. The epi-layers are subsequently etched using inductively coupled plasma reactive ion etching (ICP-RIE), with the top metal layer as an etch mask. Using Ion Beam Etching (IBE), the ground metal layer is dry-etched around the patch-antenna matrix down to the Si substrate by an Ar+ ion beam, simultaneously forming the spiral antenna or coplanar line, with the Si$_{3}$N$_{4}$ underneath preventing current leakage in the substrate. To electrically connect the patch antennas, suspended Ti/Au (20 nm/600 nm) wire bridges, approximately 150 nm wide, are fabricated through a two-step e-beam lithography process. A first resist layer serves as structural support after deposition, lithography, and reflow.  A second resist layer is then applied to define the wire bridges, followed by a standard lift-off process. This same method is used to fabricate the air-bridges connecting the 2D array to the antenna arms or the coplanar line.

%\begin{acknowledgments}
%We wish to acknowledge the support of the author %community in using
%REV\TeX{}, offering suggestions and encouragement, %testing new versions,
%\dots.
%\end{acknowledgments}

\section*{Funding} RENATECH (French Network of Major Technology Centres); Project COMPTERA - ANR 22-PEEL-0003; Contrat de Plan Etat-Region (CPER) WaveTech. Wavetech is supported by the Ministry of Higher Education and Research, the Hauts-de-France Regional council, the Lille European Metropolis (MEL), the Institute of Physics of the French National Centre for Scientific Research (CNRS) and the European Regional Development Fund (ERDF).

\section*{Acknowledgments}
We gratefully acknowledge Mathias Vanwolleghem for the carefull reading of the manuscript.

\section*{Data Availability Statement}
Data underlying the results presented in this paper are not publicly available at this time but may be obtained from the authors upon reasonable request.
\section*{References}
\nocite{*}
\bibliography{aipsamp}% Produces the bibliography via BibTeX.

\end{document}